\documentstyle[11pt]{article}

\def\mmeins#1#2#3#4#5#6#7#8#9{\mathord{1\hskip #1
    \vrule width #2 height #3 depth #4 \hskip #5
    \vrule width #6 height #7 depth #8 \hskip #9}}
\def\meins{\mathchoice
 {\mmeins{-1.50pt} {0.50pt}{7.75pt}{-0.2pt}
         {-1.20pt} {2.50pt}{0.30pt}{-0.05pt} {1.50pt}}
 {\mmeins{-1.50pt} {0.50pt}{7.75pt}{-0.2pt}
         {-1.20pt} {2.50pt}{0.30pt}{-0.05pt} {1.50pt}}
 {\mmeins{-1.17pt} {0.40pt}{5.13pt}{-0.2pt}
         {-0.67pt} {1.67pt}{0.80pt}{-0.2pt} {1.00pt}}
 {\mmeins{-1.00pt} {0.30pt}{3.85pt}{-0.2pt}
         {-0.50pt} {1.35pt}{0.80pt}{-0.2pt} {0.75pt}}}

\def\al{\alpha}

\def\gam{\gamma}
\def\Gam{\Gamma}
\def\s{\sigma}
\def\om{\omega}
\def\Om{\Omega}

\def\P{{\cal P}}
\def\Q{{\cal Q}}
\def\H{{\cal H}}
\def\I{{\cal I}}
\def\bwp{\bar{\wp}}

\def\ba{\begin{eqnarray}}
\def\ea{\end{eqnarray}}
\def\nn{\nonumber}

\begin{document}
\thispagestyle{empty}
\title{ SPECTRA OF HAMILTONIANS  WITH GENERALIZED
SINGLE-SITE DYNAMICAL DISORDER}

\author{{\sl Peter Neu}\\[0.2cm]{\it Institut f\"ur}\\
{\it Theoretische Physik}\\
{\it Universit\"at Heidelberg}\\
{\it Philosophenweg 19}\\ {\it 69120 Heidelberg,  Germany}
 \and
{\sl Roland Speicher}\\[0.2cm]
{\it Institut f\"ur }\\{\it Angewandte Mathematik}\\
 {\it Universit\"at Heidelberg}\\
{\it Im Neuenheimer Feld 294}\\{\it  69120 Heidelberg, Germany} }

\maketitle
%\newpage

\begin{abstract}
\noindent Starting from the deformed commutation relations
\ba
a_q(t) \,a_q^{\dag}(s) \  - \   q\,a_q^{\dag}(s)\,a_q(t)
 \  = \  \Gam(t-s) {\bf 1} , \quad -1\ \le \ q\ \le\ 1\nn
\ea
with
 a covariance $\Gam(t-s)$ and  a parameter $q$ varying between
$-1$ and $1$, a stochastic process  is constructed which continuously
deforms the classical Gaussian and classical  compound Poisson process.
The moments of these distinguished
 stochastic processes  are identified with the Hilbert space
 vacuum expectation values of
 products of  $\hat{\om}_q (t) = \gam\,\big(\, a_q(t) + a_q^{\dag} (t)
\,\big)\;+ \;
\xi\, a_q^{\dag} (t) a_q(t)$
 with   fixed parameters $q$, $\gam$ and $\xi$. Thereby  we can
interpolate between  dichotomic,
 random matrix and  classical
 Gaussian  and compound Poisson   processes.
 The spectra  of  Hamiltonians with single-site dynamical disorder
 are calculated for an exponential
 covariance (coloured noise) by means of the time convolution
 generalized master equation formalism (TC-GME) and the
  partial cumulants technique.
 The final result for the spectral function  is given  as a $q$-dependent
 infinite continued fraction. In the case of the random matrix
 processes the infinite continued fraction can be summed up yielding  a
 self-consistent equation for the one-particle Green function.

\newpage

\end{abstract}

\section{Introduction}
In treating the dynamics of open systems coupled to a heat
reservoir   the notion of stochasticity for the
description of the bath's influence on the small system's
dynamics is frequently introduced.
By this one means that the treatment of the
small system's dynamics is considerably simplified by
summarizing the bath's influence in some stochastic process
$\om (t)$, thus rendering the Hamiltonian itself a stochastic
process. In real physical systems  such an
approach is reliable if the temperature
is higher than the largest frequency $\om_D$
 which is present in the bath (Debye or Drude
frequency for instance).
Examples for such  situations are provided by
diffusion and relaxation phenomena in
 Brownian motion, exciton motion, laser systems, optical
line shape theory, etc.
However  the dynamics usually remains unsolvable beyond the
white noise limit \cite{HR,HS,SF}. Thus, in the past  even more simplified
models have been  considered
where the dynamical disorder is attached to one system's degree of
freedom (for instance one lattice site)  only. The Hamiltonian  then reads
\ba
H (t)  = H_0 + \om (t) V\ ,\qquad  V^2 = V = V^{\dag}
\label{hama}
\ea
where $H_0$ is the deterministic  part and  $V$ is a projector
onto the single degree of freedom to which
the stochastic process $\om (t)$ is coupled to.
This class of models  is
solvable even in the case of a bath with a coloured noise  spectrum.
Examples are the dimer problem \cite{ch1}, diffusion along
a linear chain with one point dynamical disorder \cite{ch2} or
 a random modulated trap \cite{ch3}, relaxation phenomena \cite{shibata},
etc. Approximative methods, too,  like $CPA$  reduce
 more realistic multi-site models to  single-site models and treat these
 self-consistently \cite{Lif,su1}.
The reason why single-site models are studied is that one can
 learn from these  models
about the role of stochastic processes apart from the
dichotomic (two-state-jump) and the Gaussian Markoff process  which are
commonly used, often for reasons of mathematical simplicity  only.

\noindent Here we define a new class of  stochastic processes $\hat{\om}_q(t)$
\ba
\hat{\om}_q(t) \;:=\; \gam \,\big(\,a_q(t)\,+\,a_q^{\dag} (t)\,\big)
\;+\;   \xi\, a_q^{\dag}(t) a_q(t)
\label{jjp}
\ea
 in terms of the  deformed annihilation and creation operators $a_q(t)$,
  $a_q^{\dag}(t)$.
 These operators satisfy the deformed canonical commutation relations
\ba
a_q(t) \,a_q^{\dag}(s) \  - \   q\,a^{\dag}_q(s)\,a_q(t)\  = \  \Gam(t-s)\,{\bf
1} ,
\quad -1\ \le \ q\ \le\ 1
\label{qalg}
\ea
 with  a covariance   $\Gam(t-s)$. The deformation   parameter $q$ is
  real and  varries continuously in the interval $-1\ \le \ q\ \le\ 1$.
  The limiting cases $q=1$ and $q=-1$ give the usual bosonic  and fermionic
  commutation relations on the symmetrized
  and anti-symmetrized Fock space, respectively,
  whereas the case $q=0$ allows for a realization
  on the full Fock space \cite{FB}.

 In this paper we will go beyond the white noise limit and consider
 an exponential covariance $\Gam(t-s)$  (coloured noise)
 \ba
 \Gam(t-s) := e^{-\lambda |t-s|}
 \label{expcov}
 \ea
 with modulation rate $\lambda$.
 We want to calculate the density of states (spectrum) of single-site
models (\ref{hama}) coupled to  these new kinds of stochastic processes.

\noindent Depending on $\xi$ and $q$ we can interpolate between quite
 different stochastic processes.

 The case $\xi =0$  of our process resembles a Gaussian process and will
 accordingly be named $q$-Gaussian process henceforth. It
   continuously interpolates  between
 the classical  Gaussian process ($q=1$) and
  the symmetric dichotomic process ($q=-1$) which are commonly used for
modelling
 heat reservoirs. Also the case $q=0$ gives the representation of an important
 process, namely of the Gaussian random matrix process which has relevance
 in modelling the dynamics of complex bath systems.
  In this  process the random variable
$\om (t)$ cannot be represented by a c-number function but by
an infinite, symmetric matrix whose entries are
independent  Gaussian distributed (with covariance $\Gam(t-s)$) c-number random
variables.   The probability distribution is given by
  Wigner's semicircle in the static limit.

    In the general case $\xi\ne 0$, our process and in particular its
 partial cumulants resembles a Poisson-like structure and will accordingly be
  named $q$-Poisson process.  In particular in the white noise
  limit ($\gam , \lambda , \xi \to \infty$ with
  $\gam^2/\lambda\to$ const and  $\xi/\lambda \to$ const)
   for $q=1$ we find a compound
Poisson process with an exponentially distributed jump size (see appendix).
The $(q=-1)$--Poisson process  reproduces the
biased dichotomic process, whereas the
$(q=0)$--Poisson process can again be represented by random matrices.
Namely,   the random
  variable $\om (t)$  can be chosen to be  the product of a rectangular
 $N\times M$-matrix with its adjoint in the limit
 $N,M\to\infty$ with $N/M$ fixed; the entries of this  matrix are
 again independent Gaussian  distributed c-number random variables.
 This process  will be called Poisson random matrix process  henceforth.

  By using the
time convolution generalized master equation (TC-GME)
formalism \cite{Mori}
and the partial cumulant technique \cite{pc1,pc2},
  we  calculate  exactly for both processes the spectral function of
(\ref{hama})
 for an exponential covariance (\ref{expcov}) in
form of an infinite continued fraction for all $q\in (-1,1]$.  For $q=-1$
we find a well-known finite   fraction.
 For the $q$-Poisson process  we   reproduce in the white noise limit
 the Poisson modulation model discussed by Kubo \cite{kub}.
Whereas the Gaussian noise represents spectral broadening
in a solid, i.e. scattering of a marked particle with phonons, the Poisson
noise is a model for dilute gas systems with  rare but possibly strong local
scattering events, i.e. for a bath where no harmonic modes
are present.  $\\$

\noindent Let us briefely comment on previous work.

$(1)$ The idea of deforming the canonical commutation relations
 (\ref{qalg}) in the context of stochastic processes
was first introduced  by Frisch and Bourret \cite{FB}. They
calculated the probability density  for the time-independent
commutation relations (\ref{qalg}) for all $q\in[-1,1]$.
In the time-dependent problem,
however, since the cumulant point of view was not present  at that time,
they could only treat  the special cases $q=-1$ and $q=0$ for the $q$-Gaussian
process (\ref{jjp}) properly.
For $q=0$ they rederived the Kraichnan equation \cite{krai}. \\
\indent (2) In the last time the relations (\ref{qalg}) have found increasing
interest in discussing possible deviations from the standard statistics
of identical particles \cite{Gre,wer}. Especially Greenberg proposed them as an
example
of infinite statistics interpolating between bosonic and fermionic fields.
An essential result of these investigations is the proof of the existence
of operators fulfilling (\ref{qalg}) for all $q$ with $-1\le q \le  1$
\cite{Gre,BSp,Sp2,Fiv,Zag}. This non-trivial fact allows to use (\ref{qalg})
as a sound starting point.\\
\indent (3) Mayurama and Shibata \cite{MS} interpolated between the
symmetric dichotomic and the classical Gaussian process
by composing   many symmetric dichotomic
processes $\om_n^{(s)}(t)$ to
\ba
\om^{(N)}(t) = \sum_{n=1}^{N}\om_n^{(s)}(t)\quad.
\label{aa}
\ea
This process contains as the limiting cases the dichotomic ($N=1$)
and the Gaussian ($N\to\infty$) process. The latter statement is
a consequence of the Central Limit Theorem. By using the TC-GME
and the partial cumulants  they solved the dynamics
 for the exponential covariance (\ref{expcov}) in
form of a continued fraction of length $N$.  Their powerful diagrammatic
technique
is the guiding key for the solution of our case.
By superimposing many asymmetric dichotomic processes $\om_n^{(a)}(t)$
Shibata et al. \cite{shibata,sumi}  were also the first who interpolated
between the asymmetric dichotomic process and the $(q=1)$--Poisson process.
 To our knowledge they did not consider the white noise limit in the latter
case
and therefore did not realize its  compound Poisson structure and the
connection to Kubo's Poisson modulation model.
 They named this process ``asymmetric Gaussian process".
Furthermore   since  in this approach $N$
is always finite (besides in the Gaussian limit)
 the random matrix process is not accessible. \\
\indent (4) Recently we considered the Gaussian random matrix process
by introducing a new kind of cumulants, named non-crossing cumulants
\cite{NS}.  The Gaussian random matrix process has  the Gaussian property
with respect to these cumulants,
i.e. all non-crossing cumulants higher than  second order vanish.
We derived a new generalized master-equation which is a
self-consistent  expansion  of the Green function in terms of these
cumulants and the Green function itself. For the Gaussian random
matrix process this expansion terminates after the second term
and our equation reduces to
 the  Kraichnan equation \cite{krai}. \\
\indent (5) The observation that the classical Poisson process can also be
represented with the help of creation and annihilation operators
is due to Hudson and Parthasarathy \cite{HP}. They developed a stochastic
integration theory which allows to give  a well-defined
mathematical meaning in the white noise limit to  the stochastic
Hamiltonian (\ref{hama}) itself. Their work  motivated us
to introduce the $ \xi\, a_q^{\dag}(t) a_q(t)$-term  in (\ref{jjp}).$\\$

The paper is organized as follows:  In section 2 our
processes are introduced and the partial cumulants are calculated.   In the
derivation  we will be rather condensed since the technique is by now standard
\cite{MS,sumi} and only the processes are new.
In section 3  using the TC-GME formalism together with the partial
cumulants technique
  we calculate the one-particle Green function
   exactly  for single-site disordered models
 with exponential covariance.  In section 4 we discuss our solution
 in the various limits and derive
 the corresponding spectra.
 Finally in section 5 we summarize our main results and comment
 on physical applications.
 In the appendix we recapitulate the mathematical
 structure of a classical compound Poisson process.

\section{Stochastic processes and partial cumulants}
 \noindent {\it $a)$  Moments}

It is  well-known  that a correspondence between
the Gaussian Markoff process $\om (t)$ and bosonic creation $a^{\dag}(t)$
 and annihilation operators $a(t)$ on a Hilbert space $\H$ exists. In $\H$
 the  vacuum $|0>$ is annihilated by all $a(t)$ : $a(t) |0> = 0$ for all $t$.
 We assume that  the operators fulfil the usual commutation relations
 \ba
a(t) \,a^{\dag}(s) \  - \   a^{\dag}(s)\,a(t)   = \Gamma(t-s)\  {\bf 1} ,
\label{alg1}
\ea
for a  real covariance function $\Gamma(t-s)$.
Then the vacuum expectation values of products of the operator
$\hat{\om} (t) \equiv \gam\,\big(\,a(t) + a^{\dag} (t)\,\big)$ are given by
Wick's
Theorem
\ba
<0|\hat{\om} (t_{n-1})\ldots \hat{\om} (t_1)\hat{\om} (t_0)|0> =
\left\{ \begin{array}{ll} 0 & \mbox{ $n$ odd} \\
\gam^{2m}\,\sum_{\wp_{2m}}
\Gamma(t_{i_{m-1}}-t_{j_{m-1}}) \ldots \Gamma(t_{i_0}-t_{j_0})
& n = 2m.\end{array} \right.
\label{gp}
\ea
The sum extends over all possible  partitions
$\wp_{2m}$ of the set
$\{2m-1, \ldots , 1,0\}$ into pairs $(i_{m-1},j_{m-1})$, $\ldots$, $(i_0,j_0)$
with $i_k > j_k$.
This is however exactly the formula for  the moments
$\langle \om (t_{n-1})\ldots\om (t_1)\om(t_0)\rangle$
of the Gaussian Markoff process $\om (t)$ where the brackets
$\langle\quad\rangle$ denote
the stochastic average. In this sense the operators
$\hat{\om} (t)$ give a realization of the process $\om (t)$.

In the spirit of this correspondence, we now define processes
$\om_q(t)$ for a fixed  parameter $q$ by their Hilbert space
realizations $\hat{\om}_q (t)$ in the following way:
Let $\Gamma(t-s)$
be a real covariance  and $q$  a real parameter, $-1\le q \le 1$;
let $a_q(t)$ and $a^{\dag}_q(t)$ be annihilation and
creation operators on some Hilbert space $\H$  which satisfy for
all $t$, $s$ the following deformed commutation relations:
\ba
a_q(t) \,a^{\dag}_q(s) \  - \  q\, a^{\dag}_q(s)\,a_q(t)  &=&
 \Gamma(t-s)\  {\bf 1} ,
\label{qalg2} \\
a_q(t) |0> &=& 0\ , \label{qqal}
\label{qalg3}
\ea
where ${\bf 1}$ and $|0>$ denote the identity operator and the vacuum in $\H$,
respectively.
If we now identify
\ba
\hat{\om}_q (t) := \gam\,\big(\,a_q(t) + a^{\dag}_q(t)\,\big)\;+\;
\xi\,  a^{\dag}_q(t)a_q(t)
\label{pouz}
\ea
we can define all moments of our process by calculating the vacuum
expectation values of  products of the operators $\hat{\om}_q (t)$
by means of a Wick Theorem. Using (\ref{qalg2}) and (\ref{qalg3})
this yields the following formula for the moments
\ba
\langle \om_q (t_{n-1})\ldots\om_q (t_1)\om_q(t_0)\rangle
&\equiv& <0|\hat{\om}_q(t_{n-1})\ldots
\hat{\om}_q(t_1)\hat{\om}_q (t_0)|0> \nn\\
&=& \sum_{\wp_{n}} \Gamma^{(p)}(t_{i_{p}},\ldots, t_{i_1})
\Gamma^{(q)}(t_{j_{q}},\ldots, t_{j_1})
\ldots \nn\\
&&\qquad\qquad \ldots \Gamma^{(r)}(t_{k_{r}},\ldots, t_{k_1})
\, q^{i(\wp_{n})}\ ,
\label{podef1}
\ea
where the sum extends over all possible partitions $\wp_n$ of the set
$\{n-1,\ldots,1,0\}$ into ordered
subsets  $\{i_p,\ldots,i_1\}$, $\{j_q,\ldots,j_1\}$,$\ldots$,
$\{k_r,\ldots,k_1\}$. Since we are considering a centered  process
the first moment $\langle \om_q (t_1)\rangle$ (in contrast with
the higher odd moments) vanishes
 and subsets of cardinality one do not appear,
i.e.  $p,q,\ldots, r\ge 2$.  The
$p$-point contractions  ($p>2$)
\ba
\Gamma^{(p)}(t_{i_{p}},\ldots, t_{i_1}) \; := \;\gam^2\,\xi^{p-2}\,
\Gam(t_{i_{p}}-t_{i_{p-1}})
\Gam(t_{i_{p-1}}-t_{i_{p-2}}) \ldots \Gam(t_{i_{3}}-t_{i_{2}})
\Gam(t_{i_{2}}-t_{i_{1}})\ ,
\label{mulcon}
\ea
 and  the 2-point contraction  ($p=2$)
\ba
\Gamma^{(2)}(t_1, t_0) := \gam^2 \Gam(t_{1}-t_{0})
\ea
 are expressed by  the covariance $\Gam(t - s)$ appearing  in (\ref{qalg2}).
The exponent $i(\wp_{n})$ counts
the number of  intersections  of the lines  which represent
the corresponding contractions, graphically.  Examples for this formula are
\ba
\langle  \om_q (t_{2})\om_q (t_1)\om_q(t_0)\rangle
&=& \gam^2\,\xi\; \Gam(t_2-t_1)\Gam(t_1-t_0)\\[0.25cm]
\langle \om_q (t_{3}) \om_q (t_{2})\om_q (t_1)\om_q(t_0)\rangle
&=& \gam^2\,\xi^2\Bigm(\Gam(t_3-t_2)\Gam(t_2-t_1)\Gam(t_1-t_0)\Bigm)\nn\\
&&+ \ \gam^4\Bigm( \Gam(t_3-t_2) \Gam(t_1-t_0)\; + \;
\Gam(t_3-t_0) \Gam(t_2-t_1)\nn\\
&&+ \; q\, \Gam(t_3-t_1) \Gam(t_2-t_0) \Bigm)\ .
\ea
Examples for the combinatorial exponent $i(\wp_{n})$ are

\ba
i\big(\, (3,1)\,,\,(2,0)\,\big) = 1\qquad &\doteq&\qquad
t_3\quad t_2\quad t_1\quad t_0 \nn\\[.8cm]
i\big(\, (4,2,0)\,,\,(3,1)\,\big) = 2\qquad &\doteq&\qquad
t_4\quad t_3\quad t_2\quad t_1\quad t_0 \nn\\[.8cm]
i\big(\, (5,3,1)\,,\,(4,2,0)\,\big) = 3\qquad &\doteq&\qquad
t_5\quad t_4\quad t_3\quad t_2\quad t_1\quad t_0 \nn
\ea
For the $q$-Gaussian process ($\xi\equiv 0$) all contractions (\ref{mulcon})
 with $p>2$ vanish and only pair contractions remain. Then
the definition of the
moments  simplifies to
\ba
\langle \om_q (t_{n-1})\ldots\om_q (t_1)\om_q(t_0)\rangle
\ =\   \left\{ \begin{array}{ll} 0  & \mbox{ $n$ odd} \\
\gam^{2m}\,\sum_{\wp_{2m}} \Gamma(t_{i_{m-1}}-t_{j_{m-1}})
\ldots \Gamma(t_{i_0}-t_{j_0})
\, q^{i(\wp_{2m})} & n = 2m\ .\end{array} \right.
\label{qp}
\ea
The sum extends as in (\ref{gp}) over all possible partitions
$\wp_{2m}$ of the set
$\{2m-1, \ldots , 1,0\}$ into pairs $(i_{m-1},j_{m-1}),\ldots,(i_0,j_0)$  with
 $i_k > j_k$.
This completes the definition of our processes $\om_q(t)$.

 By varying $q$  we get
a canonical  interpolation between  quite different stochastic processes.
In the special case $\xi=0$ -- the $q$-Gaussian process --
  (\ref{qp}) reduces for $q=1$ to the Gaussian
 process (\ref{gp}), for $q=-1$ we get a fermionic realization
of the dichotomic process, and for $q=0$ a realization of the
random matrix process \cite{FB,Voi,VDN,Spe1}.
In the general case  $\xi\ne0$ -- the $q$-Poisson process --
we can continuously interpolate between a compound  Poisson process ($q=1$),
the Poisson random matrix process  ($q=0$) and the asymmetric
dichotomic process ($q=-1$). $\\$

\noindent {\it $b)$ Partial cumulants}

The central quantity in the dynamical treatment in the next section
will be the partial cumulants $c^{(n)}(t_{n-1},\ldots,t_1,t_0)$
of the process (\ref{podef1}), defined as ($t_{n-1}\ge\ldots\ge t_1\ge t_0=0$)
\cite{pc1}, \cite{pc2}, \cite{MS}
\ba
c^{(n)}(t_{n-1},\ldots,t_1,t_0) :=
\langle \om_q (t_{n-1}) \Q \om_q (t_{n-2}) \Q \ldots \Q\om_q (t_1) \Q\om_q
(t_0)
\rangle\quad .
\label{defpc}
\ea
Here $\Q = \I - \P$,  and $\P$ is the averaging projector
$\P\om (t) = \langle \om (t) \rangle$, and $\I$ the identity operator
on  the  Liouville space of the small system plus the bath.
The partial cumulants are given by all irreducible
partitions $\bwp_{n}$
of the set $\{n-1, \ldots , 1,0\}$
 into ordered subsets  $\{i_p,\ldots,i_1\}$, $\{j_q,\ldots,j_1\}$,$\ldots$,
$\{k_r,\ldots,k_1\}$.
Irreducible means that the partition $\bwp_{n}$ does not split into
two disconnected partitions $\bwp^{(1)}_{n}$ and $\bwp^{(2)}_{n}$.
Thus the partial cumulants have the property of a self-energy.
Until now everything holds for a general  covariance.
For the calculation of the partial cumulants,
we now specialize to the case of an exponential covariance (coloured noise)
 (\ref{expcov}).
Then  the  diagrammatic technique introduced by Maruyama
and Shibata (\cite{MS}, Sect. 3, 4) is standard. The
partial cumulants can be  represented in form of double-staircase
diagrams  $\Lambda_{n}$ with only slightly changed rules.
 For the Gaussian case $\xi=0$,
 the contribution $c^{(n)}(\Lambda_{n})$
of all partitions $\bwp_{n}$ belonging
to such a  diagram $\Lambda_{n}$ is given by the following rules:
All upward and downward arrows $\uparrow$ , $\downarrow$
get a factor $\gam$, a downward arrow gets
 additionally the  factor $\bigm( 1 + q + \ldots q^{(k-1)}\bigm)$,
 if it connects the $k$-th
with the $(k-1)$-th level. This takes into account the different factors
$q^{i(\bwp_{2m})}$ in (\ref{podef1}). Furthermore, a horizontal
line on the $k$-th level connecting sites $i$ and $i-1$ contributes
a factor $\psi_k(t_i - t_{i-1}):= \exp(-\lambda k(t_i - t_{i-1})$.
In the general case $\xi\ne 0$ the $p$-point contractions  (\ref{mulcon}) for
$p>2$
have to be considered, additionally. As shown by Ezaki and Shibata \cite{sumi}
this only leads to a renormalization of the exponential function
$ \psi_k(t_i - t_{i-1})$. Graphically, these contractions can be considered
by inserting circles in the horizontal lines and assigning to them a factor
 $\xi\,\big(1 + q + \ldots + q^{(k-1)}\bigm)$ on the $k$-th level.
Thus we  get the following contribution
\ba
c^{(n)}(\Lambda_{n}) =
\vartheta(\Lambda_{n}) \prod_{i=1}^{n-1}
 \psi_{k_i}(t_i - t_{i-1})
\quad .
\ea
Finally, in order to get the $n$-th partial cumulant
we have to sum over all possible diagrams $\Lambda_{n}$ with
$n$ sites yielding
\ba
c^{(n)}(t_{n-1},t_{n-2},\ldots,t_1,t_0) =
 \sum_{\Lambda_{n}} c^{(n)}(\Lambda_{n}) =
 \sum_{\Lambda_{n}}
\vartheta(\Lambda_{n}) \prod_{i=1}^{n-1}
 \psi_{k_i(\Lambda_{n})}(t_i - t_{i-1})
 \label{ipo}
\ea
where $\vartheta(\Lambda_{n})$  is a $q,\gam$ and $\xi$-dependent combinatorial
factor which is determined as the product over all factors of circles and
upward and
downward arrows.  An illustrative example is
\ba
\fbox{\mbox{Diagram}}
&=& \gam\; e^{-\lambda(t_9 -t_8)}\nn\\
&&\times \ \gam\;  e^{-2\lambda(t_8 -t_7)}\;
\xi\; (1+q) \;e^{-2\lambda(t_7 -t_6)}\;\xi \;(1+q)\; e^{-2\lambda(t_6
-t_5)}\nn\\
&&\times\  \gam\; e^{-3\lambda(t_5 -t_4)}\;
 \xi \; (1+q+q^2)\; e^{-3\lambda(t_4 -t_3)}\;\nn\\
&&\times \  \gam\;   (1+q+q^2) \;e^{-2\lambda(t_3 -t_2)}\nn\\
&&\times\  \gam \;(1+q) \;
 e^{-\lambda(t_2 - t_1)}\; \xi\; e^{-\lambda(t_1 - t_0)}\;\gam\  .
\ea
The reader is referred to the original papers
\cite{MS,sumi} for more details on this technique.

\section{Dynamics}
Consider the following class of problems. Suppose the fundamental
equation of motion for some relevant system quantity $W(t)$ is
\ba
\frac{d}{dt} W(t) = -i L(t) W(t),\quad\quad W(0) = W_0\ .
\label{a}
\ea
Interpreting $W(t)$ as a wave function and $L(t)$ as the
Hamiltonian, (\ref{a}) is the Schr\"odinger equation. Similiarly,
(\ref{a})  is the Pauli master equation if $i L(t)$ is the transition rate
matrix.
Let $L(t)$ be decomposed into a deterministic, time independent part
$L_0$, characterizing the free system's dynamics, and a random, time dependent
part $L_1(t) = L_1(t,\om_q (t))$ summarizing the bath's influence
\ba
L(t) = L_0 + L_1(t,\om_q (t))\quad.
\label{b}
\ea
The free dynamics $L_0$ is supposed to be known.
We finally fix the class of models under consideration by
demanding that $L_1(t,\om_q (t))$ is  multiplicative in $\om_q (t)$
with some idempotent, deterministic system quantity (operator) $V$,
\ba
L_1(t,\om_q (t)) = \om_q (t) V,\qquad\mbox{and}\qquad V^2 = V=V^{\dag}\ .
\label{c}
\ea
It is the projector condition (\ref{c}) which allows for an exact solution
in this class of models.
Physically, it means that the dynamical disorder is attached
to one system's degree of freedom (for instance one lattice site) only.

The dynamical quantity of interest  is the one-particle Green function
(propagator) ($t > 0$)
\ba
G(t) = \left\langle \exp_{\leftarrow}{\Big \{} -i L_0t\,
-\,i\int_0^t\om_q (t') V dt'{\Big \}}\right\rangle \ ,\quad G(0) = 1
\label{d}
\ea
averaged over the stochastic process in question.
Here   $\leftarrow$ denotes the time ordering. Once
$G(t)$ is known the averaged solution of (\ref{a}) is
simply $\langle W(t) \rangle = G(t) W_0$.
The time-evolution of $G(t)$ is given in terms of its irreducible part $K(t)$
(memory function, self-energy)  by the TC-GME \cite{Mori}
\ba
\frac{d}{dt} G(t) &=& -i L_0 G(t) - \int_0^t \,K(t-t')\, G(t') \,dt' \ ,
\label{e}\\
K(t) &=& \left\langle \om_q (t) V \,\exp_{\leftarrow}{\Big \{} -i L_0t\,
-\,i\int_0^t \Q \om_q (t') V \,dt'{\Big \}}\, V \om_q (0)\right\rangle\ ,
\label{f}
\ea
where $\P$ is the averaging projector, $\P \om_q (t) =
\langle \om_q (t)\rangle$, and $\Q = \I -\P$. Since $L_0$ and $V$ are
supposed to be deterministic we have $\P L_0 = L_0\P$ and
$\P V = V \P$. Expanding (\ref{f}) the partial cumulants (\ref{defpc}) arise
\cite{CS}.
Owing to the condition (\ref{c}),  the irreducible part
 $K(t)$ has then the simple form $K(t) = k(t) V$,
 where $k(t)$ is the $c$-number function
\ba
k(t) &=& u(t) \,c^{(2)}(t,0)  \nn\\
 &&- \; i \int_0^{t} dt_1\, u(t-t_1) \, u(t_1)
 \,c^{(3)}(t,t_1,0)  \nn\\
&& - \int_0^t dt_2 \int_0^{t_2} dt_1\, u(t-t_2)\, u(t_2 - t_1)\, u(t_1)
 \,c^{(4)}(t,t_2,t_1,0)  \nn\\
 && + \; i \int_0^t dt_3 \ldots \int_0^{t_2} dt_1\,
u(t-t_3) \ldots u(t_1) \,c^{(5)}(t,t_3,t_2,t_1,0)  \nn\\
&& + \int_0^t dt_4 \ldots \int_0^{t_2} dt_1\,
u(t-t_4) \ldots u(t_1) \,c^{(6)}(t,t_4,t_3,t_2,t_1,0)  \nn\\
&&- \ldots \ \qquad .
\label{g}
\ea
Here $c^{(n)}(t,t_{n-2},\ldots,t_1,0)$ are the partial cumulants
defined in (\ref{ipo}) and
$u(t)$ is the matrix element of the free propagator
$U(t) = \exp( -iL_0 t)$ defined by $V U(t) V = u(t) V$.
The matrix element $g(t)$ of the full propagator is defined
analogously by $V G(t) V = g(t) V$.
The free
solution $u(t)$ is supposed to be known.
For the class of problems (\ref{a}) - (\ref{c})
the TC-GME (\ref{e}) can be solved exactly by means of Laplace
transformation and the usage of Dyson's equation yielding
\ba
G(z) \ =\  U(z) \ -\  U(z)\, V\, \frac{k(z)}{1 + k(z) u(z)} \,V \,U(z)\quad ,
\label{h}
\ea
where $U(z) = [z + i L_0]^{-1}$.
As it stands (\ref{h}) is exact for  processes  for which
the partial cumulants exist. We now turn to the calculation of the Laplace
transform
$k(z)$ in (\ref{h}) for our processes.

 It is now again a standard procedure to express
$k(z)$ as an infinite continued fraction \cite{MS}. The main points to notice
thereby are that on
 inserting (\ref{ipo}) in (\ref{g})   one
can pair  the  functions $u(t_i - t_{i-1})$ and
$\exp(-\lambda k(t_i - t_{i-1})$ and that one can
  define  the diagram $\Lambda_{n}$
recursively by separating the first level of the double-staircase,
$\Lambda \ =\  \uparrow\Lambda ' \downarrow \ = \
\uparrow\Lambda_1\ldots\Lambda_k \downarrow$, where the $\Lambda_i$
are the irreducible  parts of $\Lambda '$.
The circles on the horizontal lines can be  summed    in a
  geometric series.  Since the horizontal lines correspond to the free
solution,
this results in a renormalization of the free propagator
\ba
u_k^{-1}(z)\  \longrightarrow \  u_k^{-1}(z)\, +
\, i \xi q^{(k-1)}
\ea
on the $k$-th level (c.f. \cite{sumi}).
Repeating this for each
level of the double-staircase one finally succeeds in rewriting
$k(z)$ as an infinite continued fraction
\ba
k(z) = \frac{\gam^2 q^{(0)}}{\displaystyle u_1^{-1}(z) \,+\, i \xi q^{(0)} +
\frac{\gam^2 q^{(1)}}{\displaystyle u_2^{-1}(z) \,+\, i  \xi q^{(1)} +
\frac{\gam^2 q^{(2)}}{\displaystyle u_3^{-1}(z) \,+\, i \xi q^{(2)} +
\frac{\gam^2 q^{(3)}}{\displaystyle \ddots}}}}\ .
\label{kq}
\ea
with
\ba
u_k(z)&:=& u(z+k\lambda)\quad ,\\
q^{(k)}&:=& 1 + q + q^2 + \ldots + q^k = \frac{1-q^{k+1}}{1-q}\quad .
\label{qgeo}
\ea
Equations (\ref{h})  and (\ref{kq}) are the exact solution for the class
of models specified in  (\ref{a}) - (\ref{c}) for the deformed Gaussian and the
deformed Poisson process.  They are the main results in this paper.
Before we  discuss them, we want accentuate two specific cases
where the continued fraction   (\ref{kq}) can be summed
in a closed form. Namely for $q=-1$ -- {\it the dichotomic noise }
($q^{(0)} = 1$, $q^{(1)} = 0$) -- we
find  from (\ref{kq})
\ba
k(z) = \frac{\gam^2}{u_1^{-1}(z)\, +\, i\xi}\ .
\label{tt1}
\ea
For $q=0$ -- {\it the random matrix noise} ($q^{(k)} = 1$ for all $k$) --
one finds
\ba
k(z) \ =\  \frac{\gam^2}{u_1^{-1}(z) \,+\,i\xi\,+\, k(z+\lambda)}
\ =\  \frac{\gam^2}{ g^{-1}(z+\lambda)\,+\, i\xi} \quadJ.
\label{m2}
\ea
It is interesting
to note the similarity between (\ref{tt1}) and (\ref{m2}).   Loosely
speaking, one may say that the dichotomic process is the ``Born
approximation"  to the random matrix process. In the latter case
the equation (\ref{h}) is a
 non-linear self-consistent equation for $g(z)$ similiar
to those which emerge from mode-coupling  and dynamical mean field theories.

\section{Discussion}
The spectrum $\rho(\om)$ of the specified class of Hamiltonians
can be obtained
by summing the continued fraction   (\ref{kq}) numerically and
evaluating the real part of (\ref{h}) on the imaginary axis,
\ba
\rho(\om) = \frac{1}{\pi}\;\mbox{Re}\; g(z=-i\om + 0^+)\ .
\ea
It depends on the free theory through $u(z)$ and on the stochastic process
$\om_q(t)$ under consideration. Here we only discuss the latter dependence.
Setting $u^{-1}(z) = z$ (i.e. omitting $L_0$) and inserting (\ref{kq}) in
(\ref{h})  we can
rewrite the Green function  as
\ba
g(z) &= & \frac{1}{z+k(z)}\nn\\
&=&\  \frac{1}{\displaystyle z \,+ \,
\frac{\gam^2 q^{(0)}}{\displaystyle z \,+\, \lambda\, + \, i\xi q^{(0)}\,+\,
\frac{\gam^2 q^{(1)}}{\displaystyle z \,+\, 2 \lambda \, + \, i\xi q^{(1)}\,+\,
\frac{\gam^2 q^{(2)}}{\displaystyle z \,+\, 3 \lambda \, + \, i\xi q^{(2)}\,+\,
\frac{\gam^2 q^{(3)}}{\displaystyle \ddots}}}}}\ .
\label{o2}
\ea
The line shape is totally determined by the four parameters $\gam$ and $\xi$
(fluctuation strength), $\lambda$ (modulation rate or
band width of the bath) and $q$ (statistics
of the bath). The spectra can be classified into three categories:
\begin{itemize}
 \item The {\it static limit} $\gam\gg\lambda \to 0$ (for $\xi = 0$)
           and $ \gam,\xi\gg\lambda \to 0$ (for $\xi \ne 0$).
\item The {\it narrowing limit}
          $\lambda\gg\gam,\xi$ with $\gam,\xi$ fixed.
\item  The {\it white noise limit} $\gam,\lambda\to\infty$
           with $\gam^2/\lambda\to\varrho$ (for $\xi = 0$) and
		   $\gam,\xi,\lambda\to\infty$
           with $\gam^2/\lambda\to\varrho$
           and $\xi/\lambda\to\mu$ (for $\xi\ne 0$)
		   where $\varrho$ and $\mu$ are constants.
\end{itemize}
In the latter limit it is essential to scale $\xi$ with
$\lambda$ ($\xi / \lambda \to \mu$),  since otherwise
the Gaussian ($\xi = 0$) white noise limit will be found for $\xi\ne 0$
too.$\\$

\noindent{\it 1. $q$-Gaussian process --- $\xi = 0$}

{\it a) Gaussian noise:  $q=1$} -- Setting  $q=1$   provides a well-known
infinite continued fraction with $q^{(k)} = k+1$ for all $k$ in (\ref{o2})
which
can be summed exactly to
\ba
g(t)\; = \;  \exp\left[\, - \frac{\gam^2}{\lambda^2}\Bigm(
e^{-\lambda t}\,+\,\lambda t\,-\,1\Bigm)\,\right]
\label{gauq}
\ea
where $g(t)$ is the inverse Laplace transform of $g(z)$.
This is the relaxation function of a localized electron
with Gaussian energy fluctuations derived by Anderson \cite{and} and Kubo
\cite{kub2}. In the static limit one finds a Gaussian spectrum
of half-width $\gam$ around $\om=0$. In the narrowing limit
one finds a  Lorentzian at $\om=0$  whose width $\tau_{_{S}}^{-1} =
\gam^2/\lambda$
decreases with increasing modulation rate $\lambda$ (motional narrowing).
   In the white noise limit one recovers the
separation of time scales between the relaxation time of the bath
$\tau_{_B} = \lambda^{-1} \to 0$ and of the small system
$\lambda/\gam^2 \to \tau_{_S} =$ const. In both the narrowing and
the white noise limit  the dynamics shows
relaxational behaviour, i. e. an exponential  decay $\exp(-t/\tau_{_S})$.

{\it b) Symmetric dichotomic noise: $q=-1$}  --  Setting $q=-1$
terminates the continued fraction (\ref{o2}) after the second term
\ba
g(z) \;  =\;  \frac{1}{\displaystyle z + \frac{\gam^2}{z+\lambda}}\ .
\label{dico1}
\ea
In the  static limit the spectrum consists of two
$\delta$-peaks
at $\pm \gam$ which  get broadened ($\sim \lambda$) and shifted towards the
origin as $\lambda$ increases.  They describe coherent damped oscillating
behaviour.
In the narrowing and the white noise limit they
are merged into one central Lorentz peak with a width $\gam^2/\lambda$.
  The coherence is lost
and  we find the same  relaxational behaviour as for the Gaussian case.

{\it c) Gaussian random matrix noise: $q=0$} -- Setting
$q=0$  yielding $q^{(k)} = 1$ for all $k$ in (\ref{o2}) ,
one easily sums the infinite continued fraction to
\ba
g(z) \; = \; \frac{1}{z\,+\,\gam^2 g(z+\lambda)}\;=
\; u\,\Big(\,z\,+\,\gam^2 g(z+\lambda)\,\Big)
\label{rmpoi}
\ea
which is a self-consistent relation to determine $g(z)$.  Equations like
this are easily solved numerically by iteration. (It is a special
feature of this simple example that the continued fraction can be
summed exactly yielding a quotient of Bessel functions. This, however,  does
not give much practical help.) In the static limit
we can solve (\ref{rmpoi}) analytically and find Wigner's semicircle
\ba
\rho(\om)\;=\; \left\{\begin{array}{l} \frac{1}{2\pi\gam}\,\sqrt{4\gam^2-\om^2}
 \qquad  \mbox{ for $4\gam^2-\om^2\ge0$ } \\
 0\qquad \quad\qquad\quad\qquad\mbox{ for $4\gam^2-\om^2<0$ }  \ .\end{array}
\right.
 \ea
Again in the narrowing and the white noise limit this spectrum
reduces to the centered Lorentzian as in the Gaussian and the dichotomic case.
Replacing $g(z+\lambda)$
on the rhs by the free solution $u(z+\lambda) = (z+\lambda)^{-1}$
reproduces the dichotomic spectrum  (\ref{dico1}). In this sense
the random matrix process is an infinite iteration of the dichotomic process
in contrast with the Gaussian process which is the sum of infinitely many
dichotomic processes. In the language of perturbation theory,
the dichotomic process corresponds to the  Born approximation
 whereas the random matrix process represents a sum of infinitely many
 {\it non-crossing} diagrams.  Approximations like this are
 widely used in solid state physics (mode-coupling, dynamical mean-field
 theory, etc.). This latter property of the random matrix process
 has recently been exploited to define this process by a new kind of
 cumulants, the so called {\it non-crossing} cumulants \cite{NS}.
 Equation (\ref{rmpoi}) is the  Kraichnan equation \cite{krai}.

\noindent{\it 2. $q$-Poisson process --- $\xi\ne 0$. }

{\it a)  Poisson  noise: $q=1$} --   As for the Gaussian noise  the infinite
continued
 fraction (\ref{o2})   can be summed exactly to
 \ba
g(t)\; = \;  \exp\left[\, - \frac{\gam^2}{(\lambda+i\xi)^2}\Bigm(
e^{-i\xi t-\lambda t}\,+\,(\lambda + i\xi) t\,-\,1\Bigm)\,\right]\ .
\ea
First, in the static limit we  find
\ba
g(t)\ =\ e^{i\gam^2 t/\xi}\; \exp\left[\, \frac{\gam^2}{\xi^2}\,\left(
e^{-i\xi t}
- 1\right)\,\right]\ .
\ea
The spectrum   consists of
infinitely many equally spaced $\delta$-resonances
at $\om +\kappa\gam = k\xi$ with $k=0, 1, 2, \ldots$ which are
weighted according to a Poisson distribution with parameter $\kappa=\gam/\xi$
\ba
\rho(\om) \;=\; \sum_{k=0}^{\infty}\,e^{-\kappa^2}
 \frac{\kappa^{2k}}{k!}\,
\,\delta(\om+\kappa\gam-k\xi)\ .
\ea
The spectrum shows
multi-dispersive character as it is often found in physical systems
with many independent relaxation mechanisms.  \\
\indent Second, with increasing $\lambda$
and fixed $\gam$ and $\xi$ the lines
get broadened and finally  merge into one resonance at
$\om\approx -\gam^2\xi/(\lambda^2+\xi^2)$ with  width
$(\gam^2/\lambda) / ( 1+\xi^2/\lambda^2)$.
If we increase $\lambda$ further
the resonance moves toward $\om = 0$ and gets narrower.  Thus
the narrowing condition is $\lambda\gg\xi$;  then the coherence is lost.
The spectra showing transition from the static to the narrowing limit are
 depictured in Fig. 1 .\\
 \ba
 \fbox{\mbox{Fig. 1}}\nn
 \ea
  \indent Third, in the white noise  limit
 we find the relaxation function of Kubo's Poisson modulation model \cite{kub}
 with an exponentially distributed modulation strength $\al$ (remember that
 we consider centered processes)
\ba
g(t)&=& \exp\left[\,\frac{-\varrho t}{1+i\mu}\,\right]\nn\\[0.25cm]
&=& e^{i\varrho t}\,\exp\left[\,\frac{\varrho t}{\mu^2} \,
\left\langle e^{-i X} - 1\right\rangle\right]\  .
\label{relkub}
\ea
with (c.f. appendix)
\ba
dP(X = \al) \;=\; \frac{1}{\mu}\, e^{-\al/\mu}\, d\al\ , , \qquad 0\le\al <
\infty\ .
\ea
The corresponding classical process is a compound Poisson process
(see appendix).
The spectrum
has  a Lorentzian line shape at $\om = - (\mu\varrho)/(1+\mu^2)$ with  width
$\varrho/(1+\mu^2)$. Thus only for $\mu\ll 1$  (narrowing condition)
the coherence is lost. \\
\indent Kubo's Poisson modulation model consists of a modulation
\ba
M(t) = \sum_i m(t-t_i)
\ea
of independent random pulses $m(t-t_i)$ at time $t_i$ with an averaged duration
$\tau_d\to 0$, with an average  interval $\tau_i(\gg\tau_d)$ between each
pulse and an averaged height $m$.
Then $\varrho$ is identical with $\tau_i^{-1}$, i.e. $\varrho$ is the uniform
density of random pulses on the  time axis, and $\al$ is the phase shift
during the pulse \cite{kub}, i.e.
\ba
\al = \int_{-\infty}^{\infty} m(t-t_i)\, d(t-t_i)\ .
\ea
 A possible realization is $m(t-t_i) = m_i \delta(t-t_i)$.

{\it b) Asymmetric dichotomic noise: $q=-1$} --  From (\ref{o2})  we now find
\ba
g(z) \;  =\;  \frac{1}{\displaystyle z + \frac{\gam^2}{z+\lambda+i\xi}}\ .
\label{dico2}
\ea
 The $q$-Poisson process for $q=-1$
reduces to an asymmetric dichotomic process.  Thus in the static limit  the
spectrum
is compared to  the symmetric dichotomic process    shifted    to
$\om-\xi/2 = \pm\sqrt{\xi^2/4 + \gam^2}$  and the intensities of the resonance
lines  are
relatively weighted with a factor
$(1-\sqrt{1+4\gam^2/\xi^2})^2 / (1+\sqrt{1+4\gam^2/\xi^2})^2$.
In the narrowing limit we find the same behaviour as for the Poisson noise.
In the white noise limit (\ref{dico2}) reproduces  Kubo's Poisson modulation
model
(\ref{relkub}).

{\it c) Poisson random matrix noise: $q=0$} -- Again  we can sum the infinite
continued fraction (\ref{o2}) with $q^{(k)} = 1$ for all $k$ to
\ba
g(z) \; = \; \frac{1}{\displaystyle z\,+\,\frac{\gam^2
}{\displaystyle g^{-1}(z+\lambda)\,+\,i\xi}}\ ,
\label{rmpro1}
\ea
This equation can  again be solved numerically by iteration. As in the
Gaussian random matrix process we refind the dichotomic case (\ref{dico2})
on replacing $g(z+\lambda)$ by the free solution $u(z+\lambda) =
 (z+\lambda)^{-1}$. In the static limit we analytically find  the band spectrum
 \cite{Maa}
  \ba
\rho(\om)\;=\; \left\{\begin{array}{ll} \xi\,(1-\gam^2/\xi^2)\,\delta(\om\xi
+\gam^2)\; +\; \nu(\om)
 & \mbox{ for  $0\le\gam/\xi\le 1$ } \\
  \nu(\om)  & \mbox{ for $\gam/\xi >1$ }
 \end{array} \right.
 \ea
 with
 \ba
 \nu(\om) \;=\; \frac{1}{2\pi\,(\om\xi+\gam^2)}\,
\sqrt{4\gam^2 \,- \big(\om - \xi\big)^2}
 \ea
 for $\om  \in \{-\gam^2/\xi\}\cup [\xi-2\gam\, , \,
 \xi+2\gam]$ and $\rho(\om) = 0$, elsewhere.
 This function is depictured in Fig.2 a -- d for various values of $\gam /
\xi$.
 \ba
 \fbox{\mbox{Fig. 2a -- d}}\nn
 \ea
  For $\xi > \gam$ the spectrum is  a band
 of width $4\gam$ centered around $\om =\xi$  with a baggy
 semicircle line shape plus
 a $\delta$-resonance at $\om = -\gam^2/\xi$ (Fig. 2a).  The
$\delta$-resonance
 indicates  a side-band.
 With $\xi\to\gam$ the
 band and the $\delta$-resonance move towards each other and merge for $\xi =
\gam$
  (Fig. 2b); for  $\xi < \gam$
 the $\delta$-resonance has disappeared (Fig. 2c) and as $\xi\to 0$ the baggy
shape of
 spectrum vanishes and Wigner's semicircle is recovered (Fig. 2d).
 Again in  the narrowing and the white noise limit one finds
 the same behaviour as  in the previous two cases. The spectra showing the
 transition from the static to the narrowing limit are depictured in Fig. 3 .
  \ba
 \fbox{\mbox{Fig. 3}}\nn
 \ea
 \section{Summary}
 Starting from the deformed commutation relation (\ref{qalg})
  a stochastic process (\ref{jjp})  has been  constructed which continuously
deforms the classical Gaussian and classical  compound Poisson process.
 Thereby, in the case $\xi = 0$, a continuous interpolation between the
 symmetric dichotomic, the Gaussian random matrix  and the
 classical Gaussian process became possible; in the case $\xi\ne 0$,
  the interpolated processes
 were the biased dichotomic, the Poisson random matrix and the compound
 Poisson process.
To our knowledge this is  the first treatment of   such different stochastic
 processes   in one single formalism.  By coupling these processes to
one single system's degree of freedom as in (\ref{hama}) the one-particle Green
function has  been  calculated exactly
for a coloured noise bath spectrum (exponential covariance) and a general
$H_0$.

After choosing the identity operator for $H_0$
 the resulting spectra can most easily be expressed in terms of  parameters
$\gam$ and $\xi$ (fluctuation strength), $\lambda$ (modulation rate or band
width of the bath) and $q$ (statistics). They  can be classified into
three categories: $(i)$ The {\it static limit}
$\gam\gg\lambda \to 0$ (for $\xi = 0$)
   and $ \gam,\xi\gg\lambda \to 0$ (for $\xi \ne 0$).
$(ii)$ the narrowing limit $\lambda\gg\gam,\xi$ with $\gam,\xi$ fixed and
$(iii)$  the white noise limit $\gam,\lambda\to\infty$
 with $\gam^2/\lambda\to\varrho$ (for $\xi = 0$) and
$\gam,\xi,\lambda\to\infty$
 with $\gam^2/\lambda\to\varrho$
and $\xi/\lambda\to\mu$ (for $\xi \ne 0$) where $\varrho$ and $\mu$ are
constants. \\
\indent $(i)$ In the static limit  coherence with the process
is present and  the spectra strongly depend on both  the  probability
distribution (Gaussian or Poisson) and on the statistics $q$ of the process.
For the $q$-Gaussian model ($\xi = 0$)  the spectrum changes with varying $q$
from
 a Gaussian line shape ($q=1$)
  over a band   structure with  Wigner's
 semicircle law ($q=0$) to  a symmetric double-resonant line shape ($q=-1$).
 Figures illustrating this behaviour can be found in \cite{FB} and \cite{BSp}.
 For the $q$-Poisson
 model  ($\xi\ne 0$) the spectrum changes from a multi-resonant
  line shape ($q=1$)  with Poisson weighted
   intensities
 over a band  structure with  a baggy
 semicircle law  ($q=0$)
 and  a single $\delta$-resonance   ($q=-1$)
   to a biased
  double-resonant line shape. This behaviour is depictured in Fig. 4a -- e .
  \ba
  \fbox{\mbox{Fig. 4a -- e}}\nn
  \ea
\indent $(ii)$ In the narrowing limit all coherence with the attached process
is lost
 and all spectra independent  of $q$ and $\xi$
indicate exponential decay through  a Lorentzian line shape    at $\om = 0$.
The width
 $\gam^2/\lambda$  narrows with increasing modulation rate $\lambda$
 (motional narrowing).  This is understandable since in the fast
 modulation limit the small system only sees an averaged potential $\om(t)$
 which should be independent of details of the process. \\
 \indent $(iii)$  The white noise limit
 does depend on the probability distribution of the
 random variable ($\xi = 0$ or $\xi\ne 0$) but not on the statistics $q$.
 For the $q$-Gaussian model  ($\xi = 0$) we find for all $q$
  a  similiar  behaviour
as in  the narrowing limit. The coherence is lost and  all spectra
indicate exponantial decay with a time constant
 $\lambda/\gam^2\to\tau_{_S} =$ constant.
  The line shape is a Lorentzian at $\om = 0$
 with fixed width  $\tau_{_S}^{-1}$.
  For the $q$-Poisson model  ($\xi\ne 0$) we find in the white noise limit
 Kubo's Poisson modulation model with an exponentially distributed
 modulation strength for all $q$.
  The spectrum
has  a Lorentzian line shape at $\om = - (\mu\varrho)/(1+\mu^2)$ with a width
$\varrho/(1+\mu^2)$. Thus here coherence is possible and
only for $\mu\ll 1$   we find incoherent exponential relaxation. \\
 \indent  Clearly the smooth form of the spectra originates from our
simple $H_0 = \meins$  and a more sophisticated $U(z)$  in (\ref{h})
will change this picture considerably.

Physical examples where the $q$--Gaussian process has relevance are
defect systems in solids where  electrons or atoms
interact with harmonic modes (phonons) in the bath. These non-local
modes are provided by the periodic structure of the solid. Usually, in a
microscopic model, this kind of bath  is represented by an
infinite number of  harmonic oscillators with an interaction $V$ of the
structure
$V = x\sum_k(f_k^{\ast} b_k\,+\, f_k b_k^{\dag})$. Here $x$ is the defect
coordinate for instance and $b_k^{\dag}$,  $b_k$ are the creation and
annihilation operators of the oscillators which couple with a strength $f_k$,
$f_k^{\ast}$ to the defect.
Recently the $q$-Gaussian model  has been discussed in the context
of  mesoscopic conductors
with quenched disorder (static limit) \cite{mcin}; there the authors
calculated the level spacing statistics corresponding to the $q$-Gaussian
ensemble by using a random tranfer matrix formulation.

 Examples where
 the $q$--Poisson  process has physical relevance are situations where
 local, independent interactions
 are present, e. g.  dilute gas systems. Such possibly strong local
interactions
 are provided  by short-range
 scatterer which are randomly   distributed
 in space.
 Opposite to a phonon bath  no harmonic modes
 are present in  these kind of  baths.
 Such a  situation can be found in gases, fluids and amorphous solids for
example
 or generally in disordered media.

\newpage

 \section*{Acknowledgement}
 One of us (P. N.) acknowledges helpful discussions with Petr Chvosta.
 This work has been supported by grants from the
 Deutsche Forschungsgemeinschaft (R. S.) and the Graduiertenf\"orderung
 des Landes Baden-W\"urttemberg (P. N.).
 $\\$

 \section*{Figure caption}
 \noindent {\bf Fig. 1.}  Spectra of the $(q=1)$--Poisson process showing the
 transition from the multi-dispersive line shape in the static limit
 $(\lambda / \gam = 0.005)$ to the Lorentzian line shape in the narrowing
 limit   $(\lambda / \gam  = 3)$. \\

 \noindent {\bf Fig. 2a -- d.}
 Spectra of the (centered) Poisson random matrix process $(q=0)$
 in the static limit $\lambda = 0$ for different fluctuation strengths
 $\gam$ and $\xi$. For $\xi > \gam$ one finds a baggy semicircle with a single
 $\delta$-resonance ({\bf a}) which merge together for $\xi = \gam$  ({\bf b}).
 For $\xi < \gam$ the $\delta$-resonance has disappeared ({\bf c}) and the
 spectrum changes to Wigner's semicircle law as $\xi\to 0$ ({\bf d}). \\

 \noindent {\bf Fig. 3.}  Spectra of the $(q=0)$--Poisson process showing the
 transition from  a baggy semicircle line shape with a
 single $\delta$--resonance in the static limit
 $(\lambda / \gam = 0.005)$ to the Lorentzian line shape in the narrowing
 limit   $(\lambda / \gam  = 3)$. \\

 \noindent{\bf Fig. 4a -- e.} $q$--dependence of the spectrum
 of the $q$--Poisson process in the static limit $\lambda / \gam = 0.025$
and for $\xi = 2$.
  The spectrum changes from multi-dispersive behaviour (Fig. 4a -- b)
   over a band structure with a single $\delta$--resonance (Fig. 4b -- d)
 to a double-resonant form with biased intensities (Fig. 4e).

\newpage
\section*{Appendix}

\begin{appendix}

\setcounter{equation}{0}
\setcounter{section}{0}
\renewcommand{\theequation}{A\arabic{equation}}

In this appendix we show  that our $q$-Poisson process has in the white noise
limit for $q=1$ a well-defined classical interpretation as a {\it compound
Poisson process} with an exponentially distributed jump size.
Again we want to point out that the problem discussed in this paper
is trivial in the white noise limit. \\
\indent Classical white noise $\om (t)$  has the defining property that there
are no correlations
between different times, i.e. that the factorization property holds
\ba
\langle \om(t_{n-1})\ldots\om(t_{k+1}) \om(t_{k})\ldots\om(t_0)\rangle
=\langle \om(t_{n-1})\ldots\om(t_{k+1})\rangle
\langle \om(t_{k})\ldots\om(t_0)\rangle
\ea
if $\{t_{n-1},\ldots,t_{k+1}\}\cap\{t_{k},\ldots,t_{0}\} = $ \o.
Thus all correlations can be retraced to the moments $\langle\om
(t)^{n}\rangle$
which must be of the form
\ba
\langle\om (t)^{n}\rangle \;= \;c_n \,\delta^n(t)
\ea
where $\delta^n(t)$ is the $n$-dimensional $\delta$-function.
The corresponding integrated process
\ba
\Om(t) := \int_0^t \om (\tau) d\tau
\ea
has the moments (compare (\ref{podef1}))
\ba
\langle \Om (t_{n-1})\ldots\Om(t_0)\rangle
&=& \sum_{\wp_{n}} \Gamma^{(p)}(t_{i_{p}},\ldots, t_{i_1})
\Gamma^{(q)}(t_{j_{q}},\ldots, t_{j_1})
\ldots \nn\\
&&\qquad\qquad \ldots \Gamma^{(r)}(t_{k_{r}},\ldots, t_{k_1}) \ ,
\label{podef2}
\ea
with
\ba
\Gamma^{(p)}(t_{i_{p}},\ldots, t_{i_1})  \; :=\; c_p\,
\mbox{min}(t_{i_{p}},\ldots, t_{i_1}) \ .
\ea
In particular the $p$-th usual cumulant of $\Om(t)$ is  equal to $c_pt$ and the
characteristic function of $\Om (t)$ is given by
\ba
\left\langle e^{i\Om(t) x}\right\rangle\;=\; \exp\left[\,t\,\sum_{n=1}^{\infty}
\frac{c_n }{n!}(ix)^n\right] \ .
\ea
Note that our relaxation function $g(t)$ of section 4  is nothing but the
characteristic function of $\Om(t)$ evaluated at $x=-1$, i.e. in the white
noise limit
our relaxation function can be written as
\ba
g(t) = \exp\left[\,t\,\sum_{n=1}^{\infty}\frac{c_n}{n!}(-i)^n\right]\ .
\ea

\noindent Examples:
\begin{enumerate}
\item For the classical Gaussian white noise with covariance $\sigma$,
 the cumulants are given by
$c_2 = \s^2$
 and $c_n = 0$ for $n\ne 2$. The corresponding $\Om (t)$
is the {\it classical Brownian motion}  with relaxation function
\ba
g(t) = \exp\left[-\frac{\s^2 t}{2}\right]\ .
\ea
\item The classical Poisson white noise corresponds to $c_n = \varrho \al^n$
where $\al$ is the jump size and $\varrho$ the uniform density of jumps
on the time axis \cite{Fell}. The corresponding relaxation function is
\ba
g(t) = \exp\left[\, \varrho t \,\left(e^{-i\al} - 1\right)\right] \ .
\ea
\end{enumerate}
Generally, if we write $c_n = \varrho\langle X^n\rangle$ for a random variable
$X$ then the characteristic function $\Psi_X(x)$ of $X$ is equal to
\ba
\Psi_X(x) = \left\langle e^{iXx} \right\rangle =
1+\frac{1}{\varrho}\sum_{n=1}^{\infty}
\frac{c_n}{n!}(ix)^n \ .
\ea
Thus the characteristic function of the white noise $\Om(t)$  is in this case
given by
\ba
\left\langle e^{i\Om(t) x}\right\rangle\;=\;
 \exp\left[\, \varrho t \,\left(\Psi_X(x) - 1\right)\right]
\ea
and correspondingly the relaxation function by
\ba
g(t) = \exp\left[\, \varrho t \,\left(\Psi_X(-1) - 1\right)\right] \ .
\ea
In this general frame the white noise $\Om(t)$ is called a {\it compound
Poisson
process} (with respect to $X$) since it has the following realization: $\Om(t)$
behaves like a Poisson process in the sense that it jumps with a uniform
density $\varrho$.  The individual jumps
 are independent and distributed in size according to
the random variable $X$. If $X$ is not random but deterministic and takes
only one fixed value $\al$ then we recover the usual Poisson process of
example 2.

In spite of their special interpretation these compound Poisson processes are
the
most general white noises since all white noises can be recovered as limits of
compound Poisson white noises. E. g. the Brownian motion of example 1
emerges from a compound Poisson process where the random variable $X$
takes values $\pm a\s$ with equal probability and a density choosen as $\varrho
= a^{-2}$ in the limit $a\to 0$.\\
 \indent The compound Poisson process which emerges
from our $(q=1)$-Poisson process  in the white noise limit
has the cumulants
\ba
c_n = n! \frac{\varrho}{\mu^2} \mu^n = \frac{\varrho}{\mu^2}
 \left\langle X^n\right\rangle\ .
\ea
This corresponds to an exponential distribution of $X$
\ba
dP(X = \al) \;=\; \frac{1}{\mu}\, e^{-\al/\mu}\,d\al\ , \qquad 0\le\al <
\infty\ .
\ea
with the characteristic  function
\ba
\Psi_X(x) &=& \left\langle e^{iXx} \right\rangle\;=\;
\frac{1}{\mu}\int_0^{\infty}  e^{i\al x}\,e^{-\al/\mu} d\al\nn\\
&=&\frac{1}{1- ix\mu}\ .
\ea
The relaxation function therefore reads
\ba
g(t)&=& \exp\left[\,\frac{\varrho t}{\mu^2} \,
\left\langle e^{-i X} -1\right\rangle\right]\nn\\
&=&\exp\left[\,\frac{\varrho t}{\mu^2} \,
\left( \,\frac{1}{1+i\mu}\, -\,1\,\right)\right]\ .
\ea
In the main text we always consider centered processes, so that
multiplying $g(t)$ by $\exp (i\varrho t/\mu)$ reproduces our result
(\ref{relkub}).

\end{appendix}
\newpage
%\section*{References}
%\addcontentsline{toc}{section}{Literatur}

\end{document}